\documentclass[12pt]{article}
\usepackage{latexsym,amsfonts,amsmath,amsthm,amssymb}

\title{Quantum mechanics for military officers}
\author{Andrei Khrennikov\\
International Center for Mathematical Modeling\\ in Physics,
Engineering and Cognitive science\\ MSI, V\"axj\"o University,
S-35195, Sweden}

\begin{document}

\maketitle

\begin{abstract} We present a trivial probabilistic illustration
for representation of quantum mechanics as an algorithm for
approximative calculation of averages.
\end{abstract}

\section{Introduction}
In a series of authors papers \cite{KH1}--\cite{KH4} there was
developed an asymptotic approach to the problem of hidden variables
in quantum mechanics. It was shown that, in spite of all ``NO-GO''
theorems (e.g., von Neumann \cite{VN}, Cohen-Specker, Bell
\cite{B}), it is possible to construct a classical physical model,
{\it Prequantum Classical Statistical Field Theory -- PCSFT,} such
that quantum mechanics can be considered as an approximation of
PCSFT with respect to a small parameter $\alpha.$  This parameter is
given by the dispersion of fluctuations of classical random field
which is represented in quantum formalism by a von Neumann density
operator. In PCSFT the role of hidden variables is played by
classical fields. This is a model of classical statistical mechanics
with the infinite dimensional phase space. The infinite dimension
induces  a number of rather technical mathematical problems, in
particular, using theory of measures on Hilbert spaces. Such purely
mathematical difficulties embarrass understanding of PCSFT. In this
paper there will be presented a simplified version of the
approximation algorithm which induces the quantum rule for
calculation of averages -- the von Neumann trace-formula\cite{VN}.

\section{Method of Taylor approximation for functions of random
variables}

Here we follow chapter 11 of the book \cite{EV} of Elena Ventzel.
This book was the basic book for teaching probability theory in
Soviet military colleges.\footnote{I am thankful to my
farther-in-law, Alexander Choustov (marine artillery officer) who
pointed out to this chapter.} Elena Ventzel wrote her book in the
form of precise instructions what student should follow to solve a
problem:

``In practice we have very often situations in that, although
investigated function of random arguments is not strictly linear,
but it differs practically so negligibly from a linear function that
it can be approximately considered as linear. This is a consequence
of the fact that in many problems fluctuations of random variables
play the role of small deviations from the basic law. Since such
deviations are relatively small, functions which are not linear in
the whole range of variation of their arguments are {\it almost
linear} in a restricted range of their random changes,'' \cite{EV},
p. 238.

Let $y=f(x).$ Here in general $f$ is not linear, but it does not
differ so much from linear on some interval $[m_x-\delta,
m_x+\delta],$ where $x=x(\omega)$ is a random variable and
$$
m_x\equiv E \; x= \int x(\omega) \; d {\bf P}(\omega)
$$
is its average. Here $\delta > 0$ is sufficiently small. Student of
a military college  should approximate $f$ by using the first order
Taylor expansion at the point $m_x:$
\begin{equation}
\label{M} y(\omega) \approx f(m_x) + f^\prime (m_x) (x(\omega)-
m_x).
\end{equation}
By taking the average of both sides he obtains: \begin{equation}
\label{M1} m_y \approx f(m_x).
\end{equation}
The crucial point is that the linear term $f^\prime (m_x)
(x(\omega)- m_x)$ does not give any contribution! Further Elena
Ventzel pointed out \cite{EV}, p. 245: ``For some problems the above
linearization procedure may be unjustified, because the method of
linearization may be not produce a sufficiently good approximation.
In such cases to test the applicability of the linearization method
and to improve results there can be applied the method which is
based on preserving not only the linear term in the expansion of
function, but also some terms of higher orders.''

Let $y=f(x).$ Student now should preserve the first three terms in
the expansion of $f$ into the Taylor series at the point $m_x:$
\begin{equation}
\label{M2} y(\omega) \approx f(m_x) + f^\prime (m_x) (x(\omega)-
m_x) + \frac{1}{2} f^{\prime \prime} (m_x) (x(\omega)- m_x)^2.
\end{equation}
Hence
\begin{equation}
\label{M3} m_y \approx f(m_x) + \frac{\sigma_x^2}{2} f^{\prime
\prime} (m_x),
\end{equation}
where
$$
\sigma_x^2 = E \; (x - m_x)^2= \int \; (x(\omega) - m_x)^2 \; d {\bf
P}(\omega)
$$
is the dispersion of the random variable $x.$

Let us now consider the special case of symmetric fluctuations:
$$
m_x=0
$$
and let us restrict considerations to functions $f$ such that
$$
f(0)=0.
$$
Then we obtain the following special form of (\ref{M3}):
\begin{equation}
\label{M3H} m_y \approx  \frac{\sigma_x^2}{2} f^{\prime \prime} (0).
\end{equation}
We emphasize again that the first derivative does not give any
contribution into the average.

Thus at the some level of approximation we can calculate averages
not by using the Lebesgue integral (as we do in classical
probability theory), but by finding the second derivative. Such a
``calculus of probability'' would match well  with experiment. I
hope that reader has already found analogy with the quantum calculus
of probabilities. But for a better expression of this analogy we
shall consider the multi-dimensional case. Let now
$$
x=(x_1,...,x_n),
$$
so we consider a system of $n$ random variables. We consider the
vector average:
$$
m_x= (m_{x_1}, ..., m_{x_n})
$$
and the covariance matrix:
$$
B_x=(B_x^{ij}),\; B_x^{ij}= E \; (x_i -m_{x_i})\;  (x_j -m_{x_j}).
$$
We now consider the random variable $y(\omega)= f(x_1(\omega), ...,
x_n(\omega)).$ By using the Taylor expansion we would like to obtain
an algorithm for approximation of the average $m_y.$ We start
directly from the second order Taylor expansion:
$$
y(\omega) \approx f(m_{x_1},...,m_{x_n}) + \sum_{i=1}^n
\frac{\partial f}{\partial x_i}(m_{x_1},...,m_{x_n}) (x_i(\omega)-
m_{x_i})
$$
\begin{equation}
\label{Z0} + \frac{1}{2}\sum_{i,j=1}^n \frac{\partial^2 f}{\partial
x_i\partial x_j}(m_{x_1},...,m_{x_n}) (x_i(\omega)-
m_{x_i})(x_j(\omega)- m_{x_j}),
\end{equation}
and hence:
\begin{equation}
\label{Z0A} m_y \approx f(m_{x_1},...,m_{x_1})+
\frac{1}{2}\sum_{i,j=1}^n \frac{\partial^2 f}{\partial x_i\partial
x_j}(m_{x_1},...,m_{x_1}) B_x^{ij}.
\end{equation}
By using the vector notations we can rewrite the previous formulas
as:
\begin{equation}
\label{Z1} y(\omega) \approx f(m_x) + (f^\prime(m_x), x(\omega)-
m_x) + \frac{1}{2} (f^{\prime \prime} (m_x) (x(\omega)- m_x),
x(\omega)- m_x).
\end{equation}
and
\begin{equation}
\label{Z1A} m_y \approx f(m_x) + \frac{1}{2} \rm{Tr} \; B_x
f^{\prime \prime} (m_x) .
\end{equation}
Let us again consider the special case: $m_x=0$ and $f(0)=0.$ We
have: \begin{equation} \label{Z1B} m_y \approx  \frac{1}{2} \rm{Tr}
\; B_x f^{\prime \prime} (0) .
\end{equation}
We now remark that the Hessian $f^{\prime \prime} (0)$ is {\it
always a symmetric operator.} Let us now represent $f$ by its second
derivative at zero:
$$
f \to A=\frac{1}{2}f^{\prime \prime} (0).
$$
Then we see that, at some level of approximation, instead of
operation with Lebesgue integrals, one can use linear algebra:
\begin{equation}
\label{Z1BH} m_y \approx   \rm{Tr} \; B_x A
\end{equation}

\end{document}